\documentclass[prd, twocolumn, nofootinbib, floatfix]{revtex4}

\usepackage{epsfig} 
\usepackage{amsmath}

\newcommand{\beq}{\begin{equation}}
\newcommand{\eeq}{\end{equation}}
\newcommand{\beqa}{\begin{eqnarray}}
\newcommand{\eeqa}{\end{eqnarray}}
\newcommand{\om}{\Omega_m}

\newcommand{\ome}{\Omega_e}
\newcommand{\gst}{g_\star}
\newcommand{\tgst}{{\tilde g}_\star}

\newcommand{\ls}{\mathrel{\raise0.27ex\hbox{$<$}\kern-0.70em \lower0.71ex\hbox{{
$\scriptstyle \sim$}}}}

\begin{document} 

\title{Extending the Gravitational Growth Framework} 
\author{Eric V.\ Linder} 
\affiliation{Berkeley Lab \& University 
of California, Berkeley, CA 94720, USA} 
\date{\today}

\begin{abstract} 
The gravitational growth index formalism provides a model independent 
way to look for deviations from general relativity by testing dark 
energy physics distinct from its effects on the cosmic expansion history. 
Here we extend the approach to incorporate an early time parameter $\gst$ 
in addition to the growth index 
in describing the growth of large scale structure.  
We illustrate its utility for models 
with modified gravity at high redshift, early acceleration, or early dark 
energy.  Future data will have the capability to constrain 
the dark energy equation of state, the growth index $\gamma$, and $\gst$ 
simultaneously, with no degradation in the equation of state determination. 
\end{abstract} 

\maketitle

\section{Introduction \label{sec:intro}}

\vspace{-0.03in}

Exploration beyond the Standard Model of cosmology is an active 
area of investigation, trying to understand the nature of the physics 
causing the acceleration of the cosmic expansion.  For the expansion 
itself, this can be parametrized in a model independent fashion through 
the equation of state of an (effective) dark energy field.  Within 
general relativity, and for a field described entirely by its equation 
of state, this completely describes the cosmic growth history of large 
scale structure as well.  However, one should test this minimal framework. 

In general, changes in the expansion history affect the growth history.  
Rather than 
treating these as unrelated, the gravitational growth index formalism 
\cite{groexp05,lincahn07} treats in a unified fashion those effects from 
a common physical origin, and introduces a new parameter, the gravitational 
growth index $\gamma$, to keep distinct new physics such as modification 
of gravity that breaks the relation between expansion and growth.  Model 
independent parametrization of expansion, through the dark energy 
equation of state $w(a)=w_0+w_a(1-a)$, where $a$ is the cosmic scale 
factor, and of growth through $\gamma$, were merged into what was called 
Minimal Modified Gravity in \cite{hutlin07}.  Data would then be used to 
fit simultaneously the set $\{w_0,w_a,\gamma\}$.  If $\gamma$ was 
consistent with the general relativity value then no modification of 
gravity would be required. 

This framework depended on a standard matter dominated regime for 
growth at early times (see the derivation by \cite{lincahn07}) and 
could not fully treat models in which the growth was enhanced, rather 
than suppressed as expected from an accelerating component, relative 
to the matter dominated limit.  Here we broaden the formalism to 
include growth enhancement, breakdown of high redshift pure matter 
domination, and other effects of gravitational modifications. 

In Section~\ref{sec:form} we introduce the expanded formalism, Beyond 
the Standard Model 3 (BSM-3), and apply it in Section~\ref{sec:appl} to 
both enhanced and suppressed growth in various modified early universe 
scenarios.  We investigate in Section~\ref{sec:data} the ability of 
future data to constrain simultaneously the four parameters of BSM-3 
and deliver clear results on the nature of the new physics, and 
conversely the penalty in biasing results if the extended framework is 
ignored.

\section{Extending Growth \label{sec:form}} 

In seeking to understand the growth history of cosmic structure, one 
can either parametrize it directly \cite{knox,upadhye,swang,bernstein} 
or seek to use the knowledge 
that the expansion history already determines a major part of it 
(indeed all of it within the standard, general relativity scenario).  
The first path conflates the effects of expansion and growth, so we 
adopt the second approach as more likely to reveal the physics clearly. 
Also see the consistency approach of \cite{mortonson}. 
One can also explore modifications of gravity in terms of 
parametrizing the metric potential functions (see, e.g., 
\cite{caldcoo,jainzhang,zhangbean} and numerous others); this follows 
the formal application of the field equations but it is unclear how 
practical it is to relate to observations due to a number 
of unresolved issues (see the discussion in \S4.4 of \cite{linrpp}, 
and also \cite{bashinsky}).  For a general overview of modified 
gravity theories see \cite{durrer} and for classes of acceleration 
physics see \cite{uzan}. 

The gravitational growth index formalism parameterizes the growth 
itself, but takes account of the expansion history effects.  
This is more directly related to the 
observations and can be carried out practically and straightforwardly, 
but may not include all possible gravitational modifications.  In 
particular, it is not very apt at including anisotropic stress and 
other scale dependent effects.  However, many of the issues in 
\cite{linrpp,bashinsky} for the metric potential approach also concern the 
handling of anisotropic stress and scale dependence.  Certainly for linear 
growth on large 
scales the growth index formalism is a superb approximation, with 
a accuracy relative to the exact solutions in a wide range of dark 
energy and modified gravity models at the $10^{-3}$ level, and a formal 
derivation by considering deviations from the matter dominated universe 
\cite{lincahn07}. 

A first step beyond the Standard Model of cosmology with general 
relativity and a cosmological constant is to use a time varying 
dark energy equation of state to describe the expansion and growth 
histories.  This adds two parameters $w_0$ and $w_a$ to describe 
the equation of state as a function of cosmic scale factor, $w(a)= 
w_0+w_a(1-a)$, and this is generally accurate to the $10^{-3}$ level 
\cite{calib}.  We call this case with the two parameter set $\{w_0, 
w_a\}$ as Beyond the Standard Model 1, or BSM-1. 

The formalism introduced in \cite{groexp05} to test the gravitational 
framework extends this to BSM-2.  The new parameter $\gamma$, the 
gravitational growth index, was defined through 
\beq 
g(a)=e^{\int_0^a (da'/a')\,[\om(a')^\gamma-1]}, \label{eq:gnorm} 
\eeq 
where $g=(\delta\rho/\rho)/a$ is the linear growth factor of matter 
density perturbations normalized to the pure matter dominated case, 
and $\om(a)$ is the matter density fraction of the critical 
density as a function of scale factor $a$.  Thus BSM-2 involves the 
parameter set $\{w_0,w_a,\gamma\}$.  Within general relativity, 
$\gamma\approx0.55$, with a very slight variation with $w_0$, $w_a$ 
(see \cite{groexp05} for fitting forms and \cite{lincahn07} for a 
derivation). 

In the matter dominated epoch, $\om(a)\to 1$, and the presence of 
(effective) dark energy impels $\om(a)<1$ so the integrand is negative 
and growth is suppressed relative to the pure matter dominated case: 
$g<1$.  This is exactly what we would expect in the presence of an 
accelerating component.  However, one could consider particular 
circumstances where growth is enhanced.  This can formally be handled 
by allowing for a negative growth index, $\gamma<0$.  However as the 
matter dominates more completely going back to high redshifts, $\gamma$ 
is driven toward $-\infty$ if growth is for some reason enhanced.  Thus, 
the formalism gives the alarm that something non-standard is going on, 
but is not that useful in characterizing exactly what.  In addition, 
if the matter domination at high redshifts is itself non-standard, for 
example part of the nonrelativistic energy density is not due to matter, 
which clusters, but to early dark energy with an equation of state 
$w=0$, which does not cluster, then this also affects the growth index 
formalism, despite still suppressing the growth. 

Thus, although the gravitational growth index works quite well in a 
large variety of cases, there do exist exceptions that alter the 
framework in such a way that could mislead us in interpreting the 
parameter fits.  What is required is basically a calibration of the 
growth behavior at early times, how it deviates from the matter 
dominated high redshift expectation.  This is what we correct for in 
BSM-3 with the introduction of an early time calibration parameter 
$\gst$, making the simple yet palpable change:  
\beq 
g(a)=\gst\,e^{\int_0^a (da'/a')\,[\om(a')^\gamma-1]}\,. \label{eq:gstardef} 
\eeq 
We justify and discuss this form in the Appendix.  In the standard case, 
$\gst=1$.  
Enhanced growth involves $\gst>1$, and any deviation from $\gst=1$ 
signals a non-standard early universe behavior. 

So BSM-3 describes the cosmological observations at a deeper 
level with the parameter set $\{w_0,w_a,\gamma,\gst\}$.  We show 
in \S\ref{sec:data} that this enlarged set indeed delivers new insights 
while preserving the information from the expansion history at the 
constraint level of BSM-1 and most of the gravitational information 
at the level of BSM-2.  Simultaneously, however, it avoids the bias 
that would occur in BSM-1 or BSM-2 if their assumptions of the restricted 
framework were invalid. 

Note that $\gst$ serves as a necessary anchor, or calibration, for 
growth just as the parameter ${\mathcal M}$ (involving the absolute 
luminosity) is 
required for anchoring supernova distances to low redshift or 
${\mathcal S}$ (involving the absolute sound horizon scale) is 
required for anchoring 
baryon acoustic oscillation distances to high redshift (see 
\cite{linchall,linrobb,silkrecom}).  In any of the three cases one 
can avoid the need for calibration by considering only relative 
quantities, e.g.\ $d(z_1)/d(z_2)$ or $g(a_1)/g(a_2)$, but this comes 
at the price of degrading the cosmological leverage of the data; 
\cite{linrobb} showed the degradation factor is of order 2 for both 
the supernovae and baryon acoustic oscillations cases.

\section{Applications \label{sec:appl}} 

We now apply the new formalism to several specific examples to show 
that the parameters are well defined, extend the reach of the previous 
framework, and how each probes particular aspects of the physics. 

\subsection{Early Dark Energy \label{sec:ede}} 

Early dark energy refers to dark energy with a non-negligible fraction 
$\ome$ of the total energy density 
at high redshift, such as during recombination.  Models that include 
dilatation symmetries, including one of the first dark energy models 
\cite{wett88}, can have energy densities scaling at high redshift as the 
dominant component, so possessing $w=0$ and a constant fraction of 
the matter density during the matter dominated epoch.  The dark energy 
does not cluster though and the exact solution in linear growth 
theory (for $\ome\ll1$) would give $\gamma=0.6$. 

However, the equation of state of early dark energy (EDE) does not remain 
at $w=0$, but evolves toward $w=-1$ to give acceleration.  For the 
EDE parametrization of \cite{dorrobb}, the expansion history over $z\approx 
0-2$ is fit to 0.02\% by a non-EDE model with the same present equation of 
state $w_0$ and $w_a\approx 5\ome$ \cite{linrobb}.  Thus 
expansion history observations would lead us to expect a growth index 
given by the fitting formula calibrated to $10^{-3}$ accuracy on 
$w_0$-$w_a$ models \cite{groexp05} as 
\beqa 
\gamma&=&0.55+0.05\,[1+w(z=1)] \nonumber \\ 
&=&0.55+0.05\,[1+w_0+(5/2)\ome]. \label{eq:gamoe} 
\eeqa 
Note that the equation of state is supposed to be evaluated at $z=1$, 
and so one should not take $w=0$ for EDE.  For viable models, this 
gives $\gamma\approx0.55$, just like normal $w_0$-$w_a$ models.  However, 
the growth in EDE models can be very different (even from the 
$\gamma=0.6$ case) 
because the EDE impacts the high redshift matter domination. 

\begin{figure}[!htb]
\begin{center}
\psfig{file=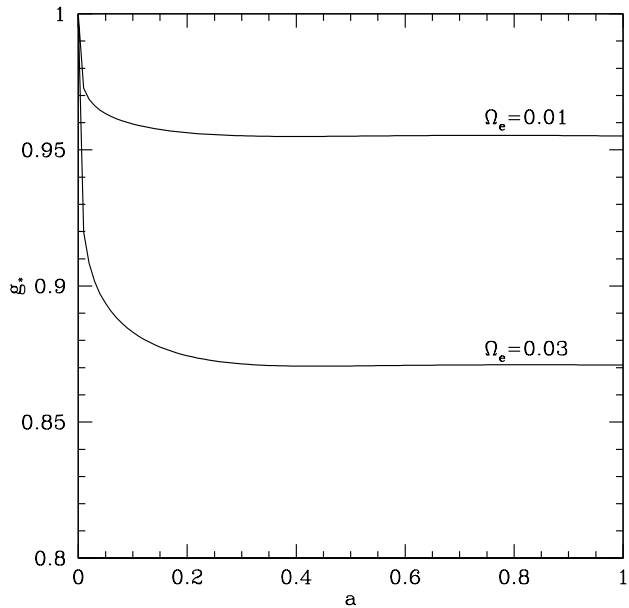,width=3.4in}
\caption{The effect of non-standard early epochs is nearly constant 
for late universe growth observations, and so it can be treated by a 
single calibration parameter $\gst$.  For the early dark energy case 
with $\ome=0.01$ (0.03), $\gst$ is constant to 0.05\% (0.2\%) for 
$z\le3$. 
}
\label{fig:edeg}
\end{center}
\end{figure}

The early time growth calibration factor $\gst$ is precisely what is 
needed to resolve the discrepancy.  Figure~\ref{fig:edeg} plots $\gst$ 
as a function of scale factor for the case $w_0=-0.95$, $\ome=0.03$.  
Over the range where growth observations can be made, $\gst$ is constant 
to high precision, 0.2\% for $z=0-3$ (1.4\% out to $a=0.1$).  This 
justifies the treatment of the calibration as a single parameter, rather 
than a function of scale factor, in the same way that the growth index 
$\gamma$ is a single, constant parameter.  Moreover, we find $\gst=0.87$, 
distinct from the standard case of $\gst=1$, giving a clear sign that 
at high redshift there is deviation from the standard matter domination 
scenario.

Because $\gst$ can be so well approximated as a constant over the 
observational epochs, this allows us to robustly define the growth 
index $\gamma$ without confusion from $\gst$.  Specifically, $\gst$ 
cancels out in the ratio $g(a_2)/g(a_1)$ for $a_1,\,a_2>0.1$.  
Fitting the growth index for this interval gives $\gamma=0.556$, in 
excellent agreement with Eq.~(\ref{eq:gamoe}) which gives 
$\gamma=0.55625$.  Indeed the deviations from the exact growth ratio 
as solved  
numerically are below 0.025\% for $z<4$ (0.11\% for $a>0.1$).  This 
approach makes clear that the growth index $\gamma$ describes the 
relative growth behavior and the new parameter $\gst$ serves as a 
calibration for the absolute growth, compared to the standard high 
redshift matter domination. 

In the EDE case, an excellent fitting form is 
\beq 
\gst=1-4.4\ome\,. \label{eq:gome} 
\eeq 
Thus a determination of $\gst$ from data leads to a constraint on 
the early dark energy density of $\sigma(\ome)\approx 0.23\,\sigma(\gst)$. 
A 10\% estimation of $\gst$ would give a tighter bound on $\ome$ than 
current constraints.  (We return to this in \S\ref{sec:data}.)  

The constancy of $\gst$ becomes even stronger for smaller $\ome$, with 
any deviations vanishing linearly with $\ome$ (e.g.\ $\delta\gst/\gst< 
0.0005$ for $z<3$ when $\ome=0.01$).  We have thus found that $\gst$ is 
truly a single parameter and one distinct from the previous (BSM-2) 
parameters, capable of extending our physics knowledge through testing 
the high redshift growth framework.

\subsection{Early Time Gravity \label{sec:grav}} 

Early universe growth will also be affected if the strength of the 
gravitational coupling was different at high redshift.  If this carries 
back to the primordial nucleosynthesis epoch, $z\approx 10^9$, then the 
variation is constrained to be less than about 10\% \cite{bbngdot}, but 
somewhat weaker limits apply around recombination, $z\approx 10^3$, about 
20\% \cite{zahnz,cmbgdot,cmbgdot2}.  

The effect of a constant multiplier $Q$ in the gravitational source term 
of the density perturbation equation is well-known (see, e.g., 
\cite{linrpp}): it delivers a growth 
\beq 
\delta\rho/\rho \sim a^{(\sqrt{1+24Q}-1)/4}\approx a^{1-(3/5)\delta Q}\,, 
\eeq 
where $\delta Q=Q-1$.  However, constant $\delta Q$ is restrictive.  
Considering the  
source term $G\om(a)$, if we leave the matter density to be given by 
the expansion history, then the gravitational modification enters through 
$G$.  By definition of the present gravitational coupling as 
$G_{\rm Newton}$ (we do not consider scale dependence here, see 
\S\ref{sec:coupl}), the variation goes to zero today so we cannot have 
a time independent $\delta Q$. 

Within scalar-tensor theories of gravity, $G$ can vary. The impact of 
this on the gravitational growth index framework was discussed in 
\cite{lincahn07} 
for a model that preserved the high redshift matter domination.  Here 
we consider the opposite situation of a deviation at early times, but 
having $G_{\rm Newton}$ today.  As a toy model we take 
\beq 
\delta Q=\frac{B}{1+(a/a_t)^q}\,. 
\eeq 
This smoothly transitions from $G=(1+B)G_{\rm Newton}$ at high redshift 
($a\ll a_t$) to $G_{\rm Newton}$ today.  When $B>0$ we expect 
enhanced growth and so $\gst>1$.  As long as $a_\star$ is not too close 
to the present, this modification is an early time phenomenon; solving 
the growth equation we indeed 
find a signal in the deviation $\gst\ne1$ while $\gamma$ is unaffected. 
Again, we see the separation of physical effects, with $\gst$ serving as 
an independent early time growth calibration parameter. 

For example, taking the expansion history to be the standard model of 
matter plus a cosmological constant, the growth within this modified 
gravity model preserves $\gamma=0.55$, but not $\gst=1$.  With $B=0.03$, 
$q=3$ (e.g.\ the scalar-tensor theory thaws from a frozen scalar field), 
and $a_t=10^{-3}$ (evolution begins near matter-radiation equality), 
we find $\gst=1.042$.  Again, describing this by a single parameter rather 
than a function of scale factor is an excellent approximation, with 
constancy preserved to 0.03\% over $a=0.1-1$.  If we choose $B<0$ then 
we have growth suppression instead, $\gst=0.959$.  The results are 
quite insensitive to the exact form of the transition in $G$, i.e.\ 
the value of $q$.  A general fitting form for $a_t>10^{-4}$ is 
\beq 
\delta\gst\approx 1.4\,B\,\log (a_t/10^{-4})\,. 
\eeq 

We emphasize that $\gamma$ is the parameter seeing deviations in the 
form of the gravitational growth equations, and $\gst$ probes 
deviations in the early time growth.  Gravitational modifications such 
as DGP braneworld theory, which look like general relativity at 
early times, give $\gst=1$.  The signal of deviation from general 
relativity for this theory appears in linear growth via the far from 
standard value of $\gamma=0.68$ \cite{luess,groexp05}.  Thus $\gst$ 
and $\gamma$ probe different aspects of the gravitational framework.

\subsection{Early Time Acceleration \label{sec:acc}} 

One of the puzzles of cosmic acceleration is the coincidence problem: 
why does acceleration happen now, within the last factor 2 in expansion 
out of the 
$\sim 10^{28}$ since inflation?  While this can be solved, or at 
least ameliorated, by use of certain time or length scales (e.g.\ 
the transition from radiation to matter domination or the magnitude 
of the scalar curvature or Hubble parameter), an alternate solution 
is removing the coincidence by having acceleration occur several times 
since inflation.  See, for example, models by \cite{griest,stoch,osc}. 

Since periods of acceleration act to suppress growth, stringent limits 
can be placed on the length of any such epochs, with \cite{darkages} 
constraining the length $\Delta\ln a$ to less than $\sim 5\%$ of the 
Hubble e-folding time based on the total growth to the present.  The 
effect on early growth due to a high redshift epoch of acceleration 
should be captured precisely by $\gst$, and furthermore we can test 
that $\gamma$ is unaffected -- i.e.\ again $\gst$ is probing distinct 
physics. 

We consider a period, beginning at $a_s$ and ending at 
$a_e=a_s\,e^{\Delta\ln a}$, when 
the dark energy density $\Omega_{\rm de}=1$ and $w=-1$, i.e.\ a 
cosmological constant-like dark energy completely dominates at 
some early epoch.  (In this toy model we take the matter density to be 
restored after this epoch to its previous value, $\om(a_e)=\om(a_s)$.)  
This indeed suppresses growth, and we find that for $z_e\gtrsim30$ 
the value of the gravitational growth index $\gamma$ for observations 
at $z\lesssim3$ is 
unaffected\footnote{Recall that for strong suppression of growth, $\gamma$ 
gets large, and due to the inertia of the growth rate, $\gamma$ does not 
fully recover until a couple of e-folds of expansion have passed.}, 
while $\gst$ is offset from unity in a manner accurately described 
by a single constant parameter.  A fitting formula is 
\beq 
\gst=1-1.2\,\Delta\ln a\,, 
\eeq 
independent of the transition time for $z_e\approx30-300$ 
(at $z>300$, the radiation effects on the growth are not completely 
negligible).  Measurements of $\gst$ could therefore directly constrain 
such epochs of early acceleration. 

Thus, we have seen that the $\gst$-$\gamma$ formalism gives an 
accurate description of 
linear growth in the presence of early dark energy density, early 
gravity, and early acceleration effects.

\subsection{Dark Coupling and Scale Dependence \label{sec:coupl}} 

While the BSM-3 framework is valuable for probing gravity and 
early time growth, it is not all encompassing.  One can have theories 
with individual features which cannot be reduced to two gravitational 
parameters.  These would be compared to observations on a theory by 
theory basis -- the point of the Beyond the Standard Model framework 
is to obtain model {\it independent\/} guidance to testing the physics. 

Two classes of theories also require further specialization.  One is 
theories involving scale dependence, where even the linear growth 
regime is not dependent purely on scale factor but behaves differently 
for different wavemodes.  An example is $f(R)$ theories (see, e.g., 
\cite{bean}).  Here a successful fitting formalism is the parametrized 
post-Friedmann approach of \cite{husawppf}.  For mild scale dependence 
due to the dark energy sound speed differing from the speed of light 
the $\gamma$ fitting formula has been extended by \cite{csgamma}.  
Another class is theories that introduce 
non-gravitational coupling between (dark) matter and dark energy. 

Dark coupling is problematic for several reasons: it 1) violates the 
Equivalence Principle, 2) introduces additional, non-gravitational 
forces, e.g.\ a non-Hubble friction term, and 3) is not well constrained 
regarding the form of the coupling.  Nevertheless, for small values 
of the coupling strength it has been shown to be well approximated 
by the $\gamma$ formalism \cite{groexp05}.  For working within the 
coupling theory class (at least for some forms of coupling), an 
accurate fitting form has been developed by \cite{amenq,diportoamen}.  

Because of remaining issues with general, rigorous treatment of 
scale dependence (see \S4.4.2 of \cite{linrpp}, and \cite{bashinsky}) 
and the ability of arbitrary forms of coupling to 
generate arbitrary growth behaviors, we do not address these classes 
further.  No finite parametrization will fit every class of model 
imaginable, so we focus on broadly applicable, model independent, 
compact yet highly accurate parametrizations.  The growth parameters 
$\gamma$ and $\gst$ serve these theoretical purposes well.  We next 
address whether they have observational practicality.

\section{Global Parameter Fits \label{sec:data}} 

While the parameter $\gst$ carries new physics insights with it, 
we have to make sure that the addition of it is practical: it must 
be reasonably constrained, and including it in the fit must not 
substantially degrade the other parameter constraints.  For example, 
the inclusion of $\gamma$ in going from BSM-1 to BSM-2 had little 
deleterious effect on the estimation of $w_0$, $w_a$.  

Because $\gst$ calibrates growth, it is degenerate with other absolute 
growth parameters such as the primordial scalar perturbation amplitude 
$A_s$ or the present equivalent, the mass fluctuation amplitude $\sigma_8$. 
Thus it will require constraints from CMB data or techniques sensitive 
to $\sigma_8$ such as weak gravitational lensing or cluster abundances 
to separate out cleanly $\gst$.  Since with present data these other 
parameters are already estimated to better than 5\%, and should improve 
further with forthcoming data, this is not an insurmountable obstacle. 

\subsection{Future Constraints \label{sec:fut}} 

To analyze the constraints on the BSM-3 parameters $\{w_0,w_a,\gamma,\gst\}$ 
we consider linear growth measurements over various 
redshift ranges, together with CMB data of Planck quality (in the form 
of a 0.2\% prior on the reduced distance to last scattering), and supernova 
distance data for $z=0-1.7$ as from a SNAP-type JDEM.  The default growth 
data has 2\% precision at $z=0.1-1$ every 0.1 in redshift, and the fiducial 
cosmology is flat $\Lambda$CDM, so $(w_0,w_a,\gamma,\gst)=(-1,0,0.55,1)$. 

\begin{figure}[!htb]
\begin{center}
\psfig{file=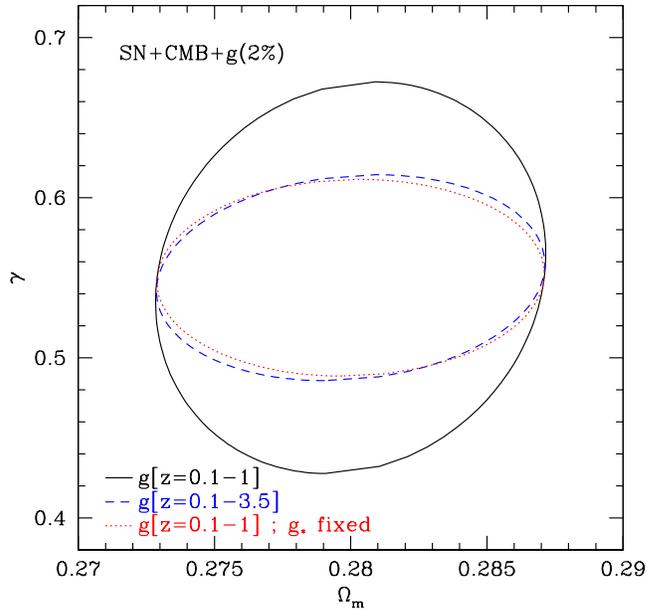,width=3.4in}
\caption{The early time calibration parameter $\gst$ plays a crucial 
role not only for physics insight, but in obtaining realistic estimates 
of the gravitational growth index $\gamma$.  The uncertainty in 
$\gamma$ when fixing $\gst$ is equivalent to assuming measurements 
extend over $z<3.5$ rather than $z<1$. 
}
\label{fig:omgam}
\end{center}
\end{figure}

The correlation coefficients between $\gst$ and the other cosmological 
parameters 
are quite small, e.g.\ $r(\om,\gst)=0.12$, with the largest one being 
$r(\gamma,\gst)=0.86$.  This lack of strong degeneracies is heartening 
and indicates that $\gst$ can function as a distinct parameter.  The 
uncertainties and full confidence contour of the equation of state parameters 
are unaffected by the addition of $\gst$ in conjunction with the growth 
data, while the fully marginalized $\sigma(\gamma)=0.081$ 
and $\sigma(\gst)=0.018$.  However, 
this does represent a factor 2.0 degradation in knowledge of $\gamma$, 
relative to fixing $\gst=1$, 
and a factor 2.0 in the area of the main growth parameters $\om$-$\gamma$ 
confidence contour, as Fig.~\ref{fig:omgam} illustrates.  This is in line 
with the other ``calibration'' parameters, where marginalizing over the 
supernova distance calibration ${\mathcal M}$ changes the expansion 
parameters $w_0$-$w_a$ contour area by a factor 1.9, and marginalizing 
over the baryon acoustic oscillation sound horizon calibration 
${\mathcal S}$ changes the $w_0$-$w_a$ contour area by 2.3.

\begin{figure}[!htb]
\begin{center}
\psfig{file=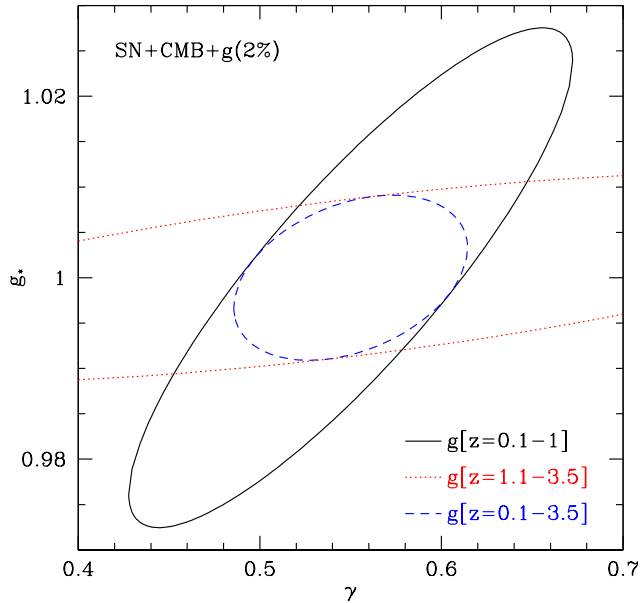,width=3.4in}
\caption{The full set of BSM-3 parameters can be fit simultaneously 
given next generation data.  The 68\% confidence level is plotted 
for the gravitational growth parameters $\gst$ and $\gamma$, marginalized 
over the expansion parameters. 
}
\label{fig:gamgst}
\end{center}
\end{figure}

Because the dependence of growth on $\gst$ is flat with redshift, 
using higher redshift data per se would not be expected to directly help 
estimation of $\gst$.  
Indeed, because sensitivity to $\gamma$ peaks at low redshift (see 
Fig.~1 of \cite{redx}), degeneracies are broken best at $z<1$.  
However, enough residual degeneracy remains that additional (not 
substitute) measurements at higher redshift do tighten constraints; 
see Fig.~\ref{fig:gamgst}. 
Extending the redshift range of growth measurements over $z=0.1-2$ 
improves the constraints more than statistically, with two times 
the number of measurements giving an almost factor two (not $\sqrt{2}$) 
improvement, to $\sigma(\gst)=0.0096$.  This carries further to 
higher redshifts, with measurements over $z=0.1-3.5$ giving a factor 
three improvement to $\sigma(\gst)=0.0060$, and $\sigma(\gamma)=0.042$. 
(Note this value of $\sigma(\gamma)$ is nearly the same as what would 
have been estimated for $z=0.1-1$ data if $\gst$ had been (improperly) 
ignored.)

Growth measurements over $z=0.1-1$ but with 1\% (4\%) precision lead 
to parameter estimation of $\sigma(\gamma)=0.042$ (0.16) and 
$\sigma(\gst)=0.011$ (0.034). 

For realistic survey design, one would have to take into account 
the observational issues in achieving the measurement precision. 
While measuring linear growth at low redshift requires surveys covering 
huge areas, the tracing objects themselves may be easier to see at 
these redshifts.  At high redshift, a broader range of scales is 
in the linear growth regime, but tracers may be more difficult to 
measure.  The Lyman-$\alpha$ forest provides a promising probe for 
the linear growth at high redshift \cite{mcdonald,lya2}, and the BOSS 
experiment \cite{boss} now underway will include a redshift range 
$z=2.3-2.8$ (in the present context the amplitude of the matter power 
spectrum rather than the baryon acoustic oscillations is the goal). 
Finally, note that all the leverage of the growth 
measurements essentially goes into the growth parameters; increasing 
the growth redshift range from $z<1$ to $z<2$ changes the area of the 
marginalized $w_0$-$w_a$ contour by 1\% (again showing that the 
new parameter is probing distinct physics). 

We could consider another form of growth measurement -- redshift 
space distortions.  This arises from a more nonlinear regime, where 
the density fluctuations induce velocities distorting the Hubble 
flow.  This was proposed as a test of the gravitational framework 
in \cite{redx,guzzo} and first analyzed in terms of BSM-2 and 
constraints on both the expansion history and gravitational growth 
index $\gamma$ \cite{redx}.  Other papers elaborating on this 
or related approaches include \cite{percivalwhite,redx3,acqua,lahav,weller}. 

The distortions $\beta$ measured through galaxy redshifts depend on 
the logarithmic derivative of the linear growth factor, and a bias 
parameter $b$ relating the galaxy density to the matter density: 
$\beta=f/b$ where 
\beqa 
f&=&\frac{d\ln g}{d\ln a}+1 \nonumber \\ 
&=&\om(a)^\gamma+\frac{d\ln\gst}{d\ln a}\,. \label{eq:fgst} 
\eeqa 
This involves the expansion history through $\om(a)$, the gravitational 
growth index $\gamma$, and the early time calibration $\gst$.  However, 
because $\gst$ is so constant, especially at late times when redshift 
distortion measurements would be evaluated, the second term in the 
last line above is negligible.  For example, for the $\ome=0.03$ 
early dark energy case where $\gst\approx0.87$ the first term 
is 100 times larger than the second. 

Adding 10\% measurements of $f$ over $z=0.1-1$ every 0.1 in redshift, 
as in \cite{redx}, to the supernova, CMB, and growth data does not 
necessarily have a significant effect on the parameter constraints.  
Treating the bias parameter $b$ as a single number and marginalizing 
over it improves the uncertainties to $\sigma(\gamma)=0.068$ and 
$\sigma(\gst)=0.016$ (i.e.\ by 15\% and 10\% respectively).  
Measurements of $f$ at the 5\% level lower these uncertainties to 0.051 
and 0.014 respectively.  These constraints weaken as a more realistic 
treatment of bias is included.  Because of this, a redshift distortion 
survey would need to be carefully designed for it to have a useful 
impact.  Note that neglecting consideration of bias or of the need 
to fit for $\gst$ could lead to overoptimistic conclusions; for example, 
ignoring $\gst$ alone makes the $\gamma$ constraint look a factor 1.8 better. 

\subsection{Parameter Bias from Ignoring $\gst$ \label{sec:bias}} 

Inclusion of the early time parameter $\gst$ not only opens new 
windows on physics but is necessary for accurate estimation of the 
gravitational growth index $\gamma$ and the expansion history parameters. 
As seen in Fig.~\ref{fig:omgam}, estimation of the uncertainty on 
$\gamma$ is significantly affected by fitting for $\gst$.  However, more 
important is that neglect of $\gst$ will bias the values of the other 
parameters. 

This can calculated within the Fisher analysis bias formalism and 
we find as a rule of thumb that $\delta\gamma\approx 3.1\,\delta\gst$. 
To preserve an accuracy of 0.05, say, in $\gamma$ to distinguish 
modified gravity from general relativity, requires that the offset 
in $\gst$ be no more than 0.016.  That is, if we neglect to fit 
$\gst$ and take by fiat $\gst=1$ (as in BSM-2), then the parameter 
estimation is substantially biased for models where the true 
$\gst<0.98$ or $\gst>1.02$. 

\begin{figure}[!htb]
\begin{center}
\psfig{file=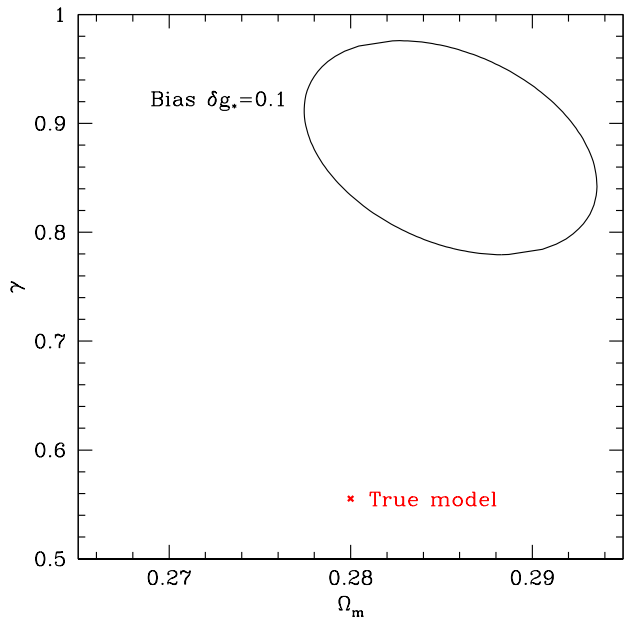,width=3.4in}
\caption{Assuming a standard $\gst=1$ rather than fitting for it 
can induce substantial bias in other cosmological parameters.  Here 
we show the offset in growth parameters (68\% cl contour) for a 
bias $\delta\gst=0.1$, as from an early dark energy model with $\ome=0.023$. 
}
\label{fig:biasomgam}
\end{center}
\end{figure}

\begin{figure}[!htb]
\begin{center}
\psfig{file=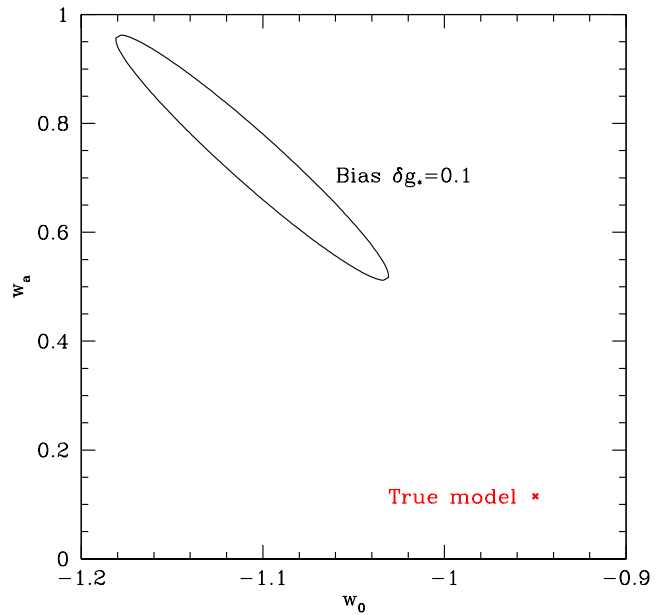,width=3.4in}
\caption{Assuming a standard $\gst=1$ rather than fitting for it
can induce substantial bias in other cosmological parameters.  Here
we show the offset in expansion parameters (68\% cl contour) 
for a bias $\delta\gst=0.1$,
as from an early dark energy model with $\ome=0.023$.
}
\label{fig:biasw0wa}
\end{center}
\end{figure}

Figures~\ref{fig:biasomgam} and \ref{fig:biasw0wa} illustrate this 
dramatically for the early dark energy case with $\Omega_e=0.023$.  
Even this small amount of early dark energy gives $\delta\gst=0.1$, 
as in Eq.~(\ref{eq:gome}). 
Through the dependence of the growth on both the gravitational growth 
index and the expansion history, all the cosmological parameters become 
substantially biased if $\gst=1$ is incorrectly assumed.  Thus, since 
we do not know a priori that early time density perturbations were 
``standard weights''\footnote{Issues of density calibration in physics 
date back to Archimedes' test of whether the king's crown was solid gold, 
and Newton's tests for counterfeit coins, but a phrase like 
``standard candles'' or even ``standard rulers'' is lacking.}, 
to protect against such bias requires simultaneously 
fitting for the early time calibration parameter $\gst$.

\section{Conclusions \label{sec:concl}} 

In exploring the nature of the physics behind the cosmic acceleration 
we can investigate both the expansion history of the universe and the 
growth history of large scale structure.  Comparison of the two is 
one of the best methods for testing our understanding of gravitation 
on large scales.  Some of the main desiderata for such a formalism are 
a compact parametrization that keeps the origins of the effects 
distinct, without conflating the physics.  Establishing a framework 
for carrying this out in a model independent manner allows us to 
search generally for the physics, including the possibility of surprises. 

Here we have extended the framework to take in important classes of 
models of both gravity and dark energy where the early time behavior 
of growth does not follow the standard matter dominated scenario.  In 
addition to the gravitational growth index $\gamma$ that reflects 
deviations in the form of the growth equations, distinct from the 
expansion history, the new parameter $\gst$ calibrates the early growth, 
distinct from $\gamma$ -- they probe different aspects of gravitation. 
This calibration is an essential element to take into account, even if 
no non-standard growth is expected, just as ${\mathcal M}$ is needed 
for robust use of supernova distances and ${\mathcal S}$ is necessary 
for robust use of baryon acoustic oscillation distances.  We have 
demonstrated that the biasing effects on other parameters if one simply 
assumes $\gst=1$ can be severe, and one will also misestimate the area 
of the growth parameters confidence contour by a factor 2. 

Future data will be capable of globally fitting the four expansion 
plus growth ``Beyond the Standard Model'' parameters 
$\{w_0,w_a,\gamma,\gst\}$.  Accurate measurements of $\gst$ can reveal 
exciting aspects of early dark energy, early gravity, and early 
acceleration, with 2\% precision delivering $\sigma(\Omega_e)=0.005$, 
or $\delta G(z\gg1)/G_{\rm Newton}$ to 1.4\%, or constraining the 
length of an early acceleration period to 1.7\% of a Hubble time. 

These prospects are exciting.  To achieve them will require good 
measurements of the degenerate parameters of the primordial or current 
density perturbation amplitude, $A_s$ or $\sigma_8$, and linear 
growth measurements, perhaps through the Lyman-$\alpha$ forest, of 
$\sim 2\%$.  The inclusion of $\gst$ necessarily increases the uncertainty 
on $\gamma$ by a factor 2.  To recover the precision will require a 
very broad redshift range $z\approx0-3.5$ of linear growth measurements. 
Alternately, weak gravitational lensing data may help, though we have 
seen that redshift distortions do not give substantial tightening, and 
nonlinear structure may not 
either \cite{fllaug08,grossi} (but see \cite{georgcoupl}). 

The BSM-3 framework delivers extended physics information over a 
broader class of models, including those with enhanced growth or 
breaking standard matter domination.  Adding a single early time 
parameter $\gst$ is a highly accurate approximation, constant over 
the observable epoch to the $10^{-3}$ level and agreeing with the 
exact growth solution at the $10^{-3}$ level.  Moreover, the framework is 
practical, able to constrain deviations from general relativity and 
the standard model while keeping the physical interpretations clear 
and distinct.

\section*{Appendix: Early time treatment \label{sec:apx}} 

Rather than the approach adopted in this article, one might have 
considered defining a parameter $\tgst$ so as to represent the 
{\it entire\/} early growth history -- that is, 
\beq 
g(a)=\tgst\,e^{\int_{a_*}^a (da'/a')\,[\om(a')^\gamma-1]} \,, 
\eeq 
in contrast to the lower limit of 0 for the integral in 
Eq.~(\ref{eq:gstardef}). 
However, this conflates the expansion history physics with 
non-standard growth physics.  For example, for a model with 
$(w_0,w_a)=(-0.8,0.5)$ and standard gravity, one would derive 
$\tgst=0.95$, apparently indicating a non-standard high 
redshift framework.  In other words, using the definition of 
$\tgst$, rather than $\gst$, one has introduced a parameter that 
is not a distinct 
probe from $\gamma$ and the expansion history $\om(a)$, possibly 
leading to confusion.  By keeping the full evolution of the $\gamma$ 
term and defining $\gst$ as in Eq.~(\ref{eq:gstardef}), one clearly 
separates the different physical effects beyond standard cosmology.

\acknowledgments 

I thank the Michigan Center for Theoretical Physics for 
hospitality during part of this work.  
This work has been supported in part by the Director, Office of Science, 
Office of High Energy Physics, of the U.S.\ Department of Energy under 
Contract No.\ DE-AC02-05CH11231.

\end{document}